\documentclass[aps,preprint,nofootinbib]{revtex4-1}
\usepackage{graphicx}
\usepackage{epstopdf}
\usepackage{amsmath}
\usepackage{amssymb}

\usepackage{multirow}
\usepackage{color}

\usepackage[greek,english]{babel}
\languageattribute{greek}{polutoniko}
\usepackage[iso-8859-7]{inputenc}

\usepackage{tikz}
\usetikzlibrary{shapes.geometric, arrows}

\tikzstyle{startstop} = [rectangle, rounded corners, minimum width=3cm, minimum height=1cm,text centered,text width=\columnwidth/2, draw=black]
\tikzstyle{io} = [trapezium, trapezium left angle=70, trapezium right angle=110, minimum width=3cm, minimum height=1cm,text width=\columnwidth/3, text centered, draw=black]
\tikzstyle{process} = [rectangle, minimum width=3cm, minimum height=1cm, text centered, text width=\columnwidth/3, draw=black]
\tikzstyle{decision} = [diamond, minimum width=1cm, minimum height=1cm,text centered, text width=\columnwidth/6,draw=black]
\tikzstyle{arrow} = [thick,->,>=stealth]
\tikzstyle{start} = [ellipse, minimum width=1cm, minimum height=1cm,text centered,text width=\columnwidth/3, draw=black]

\def\dalemb#1#2{{\vbox{\hrule height.#2pt
  \hbox{\vrule width.#2pt height#1pt \kern#1pt \vrule width.#2pt}
    \hrule height.#2pt}}}

\def\ba{\begin{eqnarray}}
\def\ea{\end{eqnarray}}
\def\be{\begin{equation}}
\def\ee{\end{equation}}

\def\gtorder{\mathrel{\raise.3ex\hbox{$>$}\mkern-14mu
             \lower0.6ex\hbox{$\sim$}}}
\def\ltorder{\mathrel{\raise.3ex\hbox{$<$}\mkern-14mu
             \lower0.6ex\hbox{$\sim$}}}

\begin{document}


\preprint{T$\Theta\Delta$}

\title{Integrated probability of coronary heart disease 
subject to the --308 tumor necrosis factor--$\alpha$ SNP: a Bayesian meta--analysis
}

\author{
C. Sofia~Carvalho${}^{1,2}$\footnote{Email: cscarvalho@oal.ul.pt}
}
\address{$^1$ 
Institute of Astrophysics and Space Sciences, University of Lisbon,
OAL, Tapada da Ajuda, 1349-018 Lisboa, Portugal}
\address{$^2$
Research Center for Astronomy and Applied Mathematics, Academy of Athens, Soranou Efessiou 4, 11-527, Athens, Greece}
\begin{abstract}

We present a meta--analysis of independent studies on the potential implication in the occurrence of coronary heart disease (\text{CHD}) of the single--nucleotide polymorphism (SNP) at the --308 position of the tumor necrosis factor alpha (TNF--$\alpha$) gene. 
We use Bayesian analysis to integrate independent data sets and to infer statistically robust measurements of correlation. Bayesian hypothesis testing indicates that there is no preference for the hypothesis that the --308 TNF--$\alpha$  SNP is related to the occurrence of \text{CHD,} in the Caucasian or in the Asian population, over the null hypothesis. As a measure of correlation, we use the probability of occurrence of CHD conditional on the presence of the SNP, derived as the posterior probability of the Bayesian meta--analysis. The conditional probability indicates that CHD is not more likely to occur when the SNP is present, which suggests that the --308 TNF--$\alpha$ SNP is not implicated in the occurrence of CHD.
\end{abstract}




\date{\today}

\pacs{}

\maketitle

\section{Introduction}

Coronary heart disease (\text{CHD}) is now widely accepted to consist of a chronic inflammatory disease \cite{hansson_2005}.
CHD is a complex disease with multifold etiology, with both genetic and environmental factors contributing to its occurrence and development. 

Among the genetic factors potentially implicated in the emergence of \text{CHD,} the tumor necrosis factor alpha (TNF--$\alpha$) has attracted a great interest for its involvement in the inflammatory response of the immune system \cite{vassali_1992}.
There is evidence that TNF-$\alpha$ is implicated in an increased susceptibility to the pathogenesis of a variety of diseases. 
In particular, high serum levels of TNF--$\alpha$ affect endothelial cell hemostatic function and hence may modify the risk for developing CHD \cite{plutzky_2001}.
There is also the suggestion that the TNF--$\alpha$ gene affects the modulation of lipid metabolism, obesity susceptibility and insulin resistance, 
thus being potentially implicated in the development of CHD (see Ref.~\cite{vourvouhaki_2008} and references therein).

Among the several single--nucleotide polymorphisms (SNPs) that have been identified in the human TNF-$\alpha,$ the best documented one is at the position --308 of the TNF--$\alpha$ gene promoter. This SNP involves the substitution of guanine (G) for adenine (A) and the subsequent creation of two alleles (TNF1(A) and TNF2(G)) and three genotypes (GG, GA and AA) \cite{wilson_1992}.
It has been hypothesised that the TNF--$\alpha$ SNP could change the susceptibility to CHD. However, the results on its association with CHD are contradictory, some implying different influence of the two alleles on the prevalence of \text{CHD,} others implying no association (see Ref.~\cite{zhang_2011} and references therein).

In order to infer the risk of CHD derived from potential risk factors, it is important to develop a formalism that infers correlations among different intervening factors and combines independent data sets for a consistent inference of the correlations. In Ref.~\cite{vourvouhaki_2009} we introduced a formalism based on Bayesian inference to infer the correlation of the occurrence of CHD with two risk factors and tested a simplistic model for the signal pathway on the three--variable data set from Ref.~\cite{vendrell_2003}. 
In this manuscript we extend the formalism to extract information from the combination of data from independent studies and to quantify the combined risk of occurrence of CHD from the --308 TNF--$\alpha$ SNP. 

The most exhaustive meta--analysis to date on this correlation is the frequentist analysis in Ref.~\cite{zhang_2011} covering Caucasian, Asian, Indian and African populations. This meta--analysis found a 1.5 fold increased risk of developing CHD when the SNP is present in the Caucasian population, but found no association in the other ethnicities. A more recent meta--analysis, covering the same data sets, found no association in the Caucasian or in the Asian population \cite{chu_2013}.

\begin{figure}[h]
\centering
\begin{tikzpicture}[node distance=2.cm]
\node (start1) [start] {DATA SELECTION};
\node (in1) [startstop, below of=start1,yshift=0.6cm] {Select studies};
\node (pro1a) [process, right of=in1,xshift=\columnwidth/3] {Compute fraction of SNP in population of CHD patients, $f_{\text{SNPinCHD}}$};
\node (pro1b) [process, below of=pro1a,yshift=-0.3cm] {Compute fraction of SNP in population of non--CHD patients, $f_{\text{SNPin}\overline{\text{CHD}}}$};
\node (dec1) [startstop, below of=pro1b] {Ratio $f_{\text{SNPinCHD}}/f_{\text{SNPin}\overline{\text{CHD}}}$ indicates correlation sign};
\draw [arrow] (in1) -- (pro1a);
\draw [arrow] (pro1a) -- (pro1b);
\draw [arrow] (pro1b) -- (dec1);
\end{tikzpicture}
\vskip0.2cm
\centering
\begin{tikzpicture}[node distance=2.cm]
\node (start2) [start] {HYPOTHESES TESTING};
\node (pro2a) [process, below of=start2,yshift=0.2cm]{Define hypothesis $H_{0}:$ presence of SNP is unrelated to occurrence of CHD};
\node (pro2b) [process, below of=pro2a,yshift=-0.3cm]{Define hypothesis $H_{1}:$ presence of SNP is related to occurrence of CHD};
\node (pro2c) [process, below of=pro2b,yshift=-0.3cm] {Compute evidence of each hypothesis, $P(D_{\text{\text{SNP}}}\vert H_{0})$ and $P(D_{\text{\text{SNP}}}\vert H_{1})$};
\node (dec2) [startstop, below of=pro2b, yshift=-2.3cm] {Ratio of evidence $B_{10}$ 
indicates which hypothesis is favoured by data};
\node (pro2d) [process, right of=dec2,xshift=\columnwidth/3]{Compute $B_{10}$ as function of ratio $f_{\text{SNPinCHD}}/f_{\text{SNPin}\overline{\text{CHD}}}$ to complement hypothesis testing with correlation sign};
\draw [arrow] (pro2a) -- (pro2b);
\draw [arrow] (pro2b) -- (pro2c);
\draw [arrow] (pro2c) -- (dec2);
\draw [arrow] (dec2) -- (pro2d);
\end{tikzpicture}
\caption{\baselineskip=0.5cm{
{\bf Flow Chart.} Panel 1 of 3. Ellipses indicate the main actions. Rectangles indicate detailed actions. Rectangles with rounded corners indicate the main results.}}
\label{fig:flowchart_1}
\end{figure}
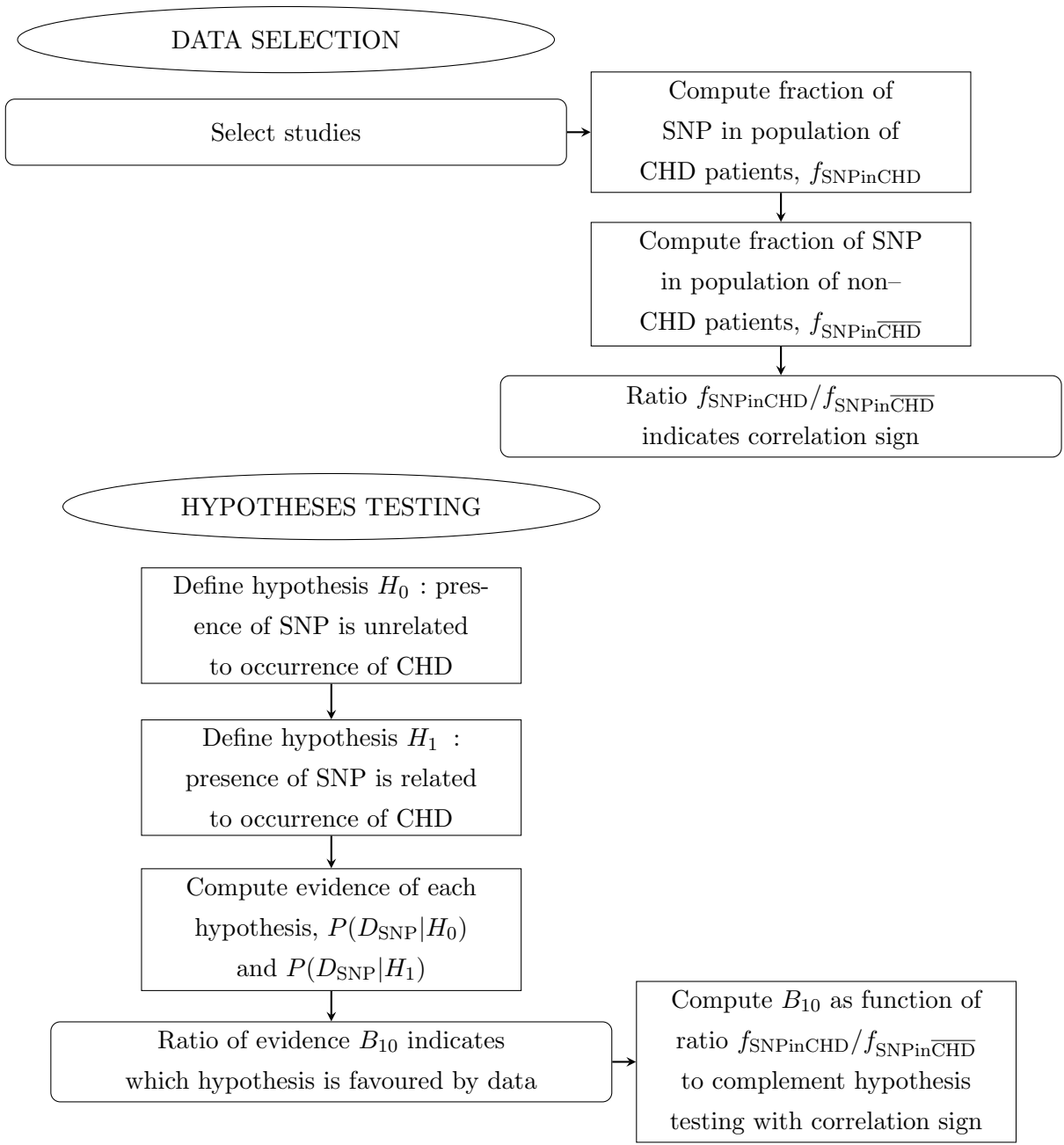

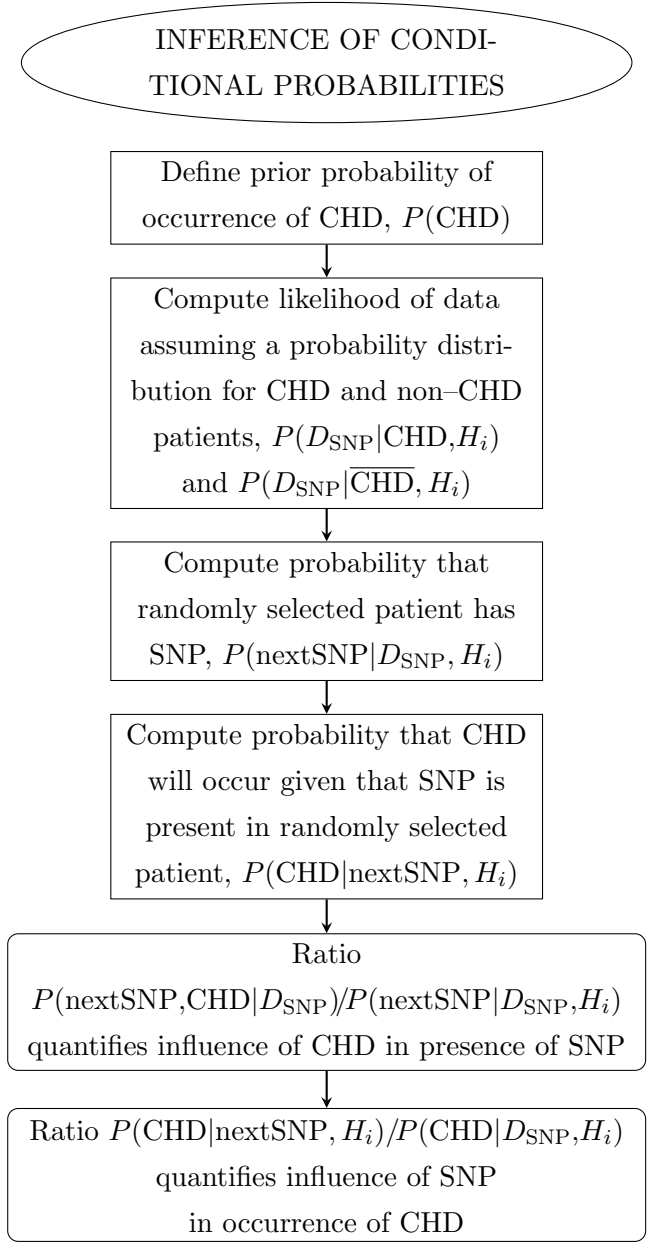
\begin{figure}[t]
\centering
\begin{tikzpicture}[node distance=2.cm]
\node (start3) [start]{INFERENCE OF CONDITIONAL PROBABILITIES};
\node (pro3a) [process, below of=start3,yshift=0.2cm]{Define prior probability of occurrence of CHD, $P(\text{CHD})$};
\node (pro3b) [process, below of=pro3a,yshift=-0.6cm]{Compute likelihood of data assuming a probability distribution for CHD and non--CHD patients, $P(D_{\text{SNP}}\vert \text{CHD,} H_{i})$ and $P(D_{\text{\text{SNP}}}\vert \overline{\text{CHD}}, H_{i})$};
\node (pro3c) [process, below of=pro3b,yshift=-0.9cm]{Compute probability that randomly selected patient has SNP, $P(\text{nextSNP}\vert D_{\text{SNP}}, H_{i})$}; 
\node (pro3d) [process, below of=pro3c,yshift=-0.6cm]{
Compute probability that CHD will occur given that SNP is present in randomly selected patient, $P(\text{CHD}\vert \text{nextSNP}, H_{i})$};
\node (dec3a) [startstop, below of=pro3d, yshift=-0.6cm] {Ratio 
  $P(\text{nextSNP},\!\text{CHD} \vert D_{\text{\text{SNP}}})\!/\!P(\text{nextSNP}\vert D_{\text{\text{SNP}}},\!H_{i})$ 
quantifies influence of CHD in presence of SNP};
\node (dec3b) [startstop, below of=dec3a, yshift=-0.3cm] {Ratio
$P(\text{CHD}\vert \text{nextSNP},H_{i})/\!P(\text{CHD}\vert D_{\text{SNP}},\!H_{i})$ 
quantifies influence of SNP in occurrence of CHD};
\draw [arrow] (pro3a) -- (pro3b);
\draw [arrow] (pro3b) -- (pro3c);
\draw [arrow] (pro3c) -- (pro3d);
\draw [arrow] (pro3d) -- (dec3a);
\draw [arrow] (dec3a) -- (dec3b);
\end{tikzpicture}
\caption{\baselineskip=0.5cm{
{\bf Flow Chart.} Panel 2 of 3.}}
\label{fig:flowchart_2}
\end{figure}

\begin{figure}[h]
\vskip0.4cm
\centering
\begin{tikzpicture}[node distance=2.cm]
\node (start4) [start]{SENSITIVITY OF RESULTS};
\node (pro4) [process, below of=start4,yshift=-0.1cm]{Repeat calculations by excluding:
a) low--significant data sets + data with extreme results; b) extreme data sets};
\node (dec4) [startstop, below of=pro4, yshift=-0.2cm]{Proof of formalism's robustness};
\draw [arrow] (pro4) -- (dec4);
\end{tikzpicture}
\caption{\baselineskip=0.5cm{
{\bf Flow Chart.} Panel 3 of 3.}}
\label{fig:flowchart_3}
\end{figure}
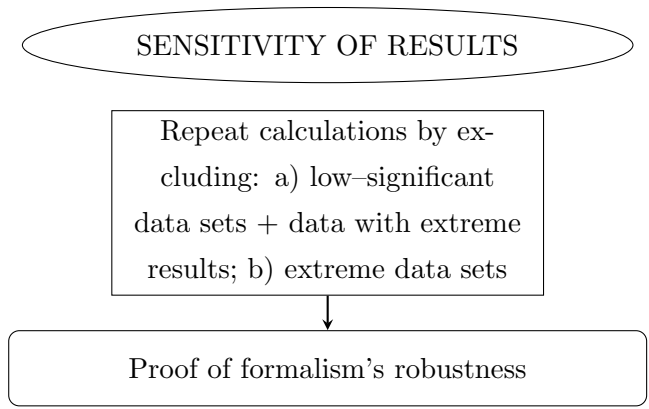

In this manuscript we propose a meta--analysis based on Bayesian analysis in an attempt to establish the potential implication of --308 TNF--$\alpha$ SNP in the occurrence of CHD.
This manuscript is organized as follows. In Section \ref{sec:methods} we describe the method. 
In particular, in Subsection \ref{sec:data} we describe the data sets selected; in Subsection \ref{sec:hypo} we propose two hypotheses and test which best and most simply describes the data.
In Section \ref{sec:results} we perform the Bayesian analysis of the selected data sets, combined by ethnicity and CHD phenotype, and present the results.
In particular, in Subsection \ref{sec:infer} we infer the conditional probabilities for the occurrence of CHD given the presence of the SNP;
in Subsection \ref{sec:sens} we test the sensitivity of this formalism to low--significance data sets, to data sets with extreme results and to extreme data sets. 
Finally in Section \ref{sec:concl} we draw the conclusions.
Below there follows a flow chart describing summarily the reasoning of this meta--analysis (Fig.s~\ref{fig:flowchart_1},\ref{fig:flowchart_2}, and \ref{fig:flowchart_3}).

\section{Methods}
\label{sec:methods}

\subsection{Data Selection}
\label{sec:data}

\begin{table}[t]
\begin{tabular}{c|c|cc|cc|cc}
\hline
Study & ~Phenotype~ 
& \multicolumn{2}{c|}{CHD patients} 
& \multicolumn{2}{c|}{Controls}
&  \multicolumn{2}{c}{Bayes factor} 
 \\ 
($i$) & ($j$) 
~&~ GG ~&~ GA/AA 
~&~ GG ~&~ GA/AA 
~&~$(H^{i,j}_{1}/H^{i,j}_{0})$~ &~$(H^{j}_{1}/H^{j}_{0})$\\ 
\hline\hline
Allen et al. (NA) &\multirow{6}{*}{Cauc CS}&127 &53 &222 &107 &$0.14\pm0.05$ 
&\multirow{3}{*}{$0.049\pm 0.014$}\\
Elahi et al. (A) &&59 &38 &41 &54 &$3.54\pm 1.12$&\\
Georges et al. (A) &&613 &236 &222 &92 &$0.08\pm 0.03$&\multirow{2}{*}{$0.041\pm 0.016^{\ast}$}\\
Sbarsi et al. (A) &&175 &73 &185 &56 &$0.33\pm 0.11$&\\
Szalai et al. (A) &&229 &89 &181 &87 &$0.19\pm 0.07$&\multirow{1}{*}{$0.048\pm 0.019^{\ast\ast}$}\\
Vendrell et al. (A) &&231 &110 &159 &48 &$1.33\pm 0.46$&\\
\hline
Antonicelli (A) &\multirow{9}{*}{Cauc MI} &224 &69 &246 &64 &$0.12\pm 0.04$
&\multirow{6}{*}{$0.026\pm 0.011$}\\ 
Bennet et al. (A) &&799 &368 &1037 &460 &$0.05\pm 0.02$&\\  
Dedoussis et al. (A) & &206 &31 &227 &10 &$26.14\pm 8.56$&\\
Herrmann et al.$^{\dagger}$ (NA) &&325 &120 &376 & 158 &$0.11\pm 0.04$&\multirow{3}{*}{$0.035\pm 0.015^{\ast}$}\\ 
Herrmann et al.$^{\ddagger}$ (NA) &&117 &79 &97 & 79 &$0.19\pm 0.06$&\\ 
Koch et al. (NA) &&565 &228 &244 &96 &$0.07\pm 0.03$&\multirow{2}{*}{$0.030\pm 0.012^{\ast\ast}$}\\ 
Padovani et al. (A) &&120&28&114&34&$0.17\pm 0.06$&\\
Tobin et al. (A) &&365 &182 &337 &168 &$0.07\pm 0.03$&\\
Tulyakova et al. (NA) &&242 &64 &177 & 69 &$0.60\pm 0.21$&\\
\hline
Chen et al. (NA) &\multirow{5}{*}{Asian CS} &29 &11 &21 &9 &$0.27\pm0.08$ 
&\multirow{2}{*}{$0.151\pm 0.057$}\\
Hou et al. (NA) && 268 &32 &802 &103 &$0.05\pm0.02$ &\\
Li et al. (NA) && 66 &8 &138 &20 &$0.12\pm0.04$& \multirow{1}{*}{$0.114\pm 0.043^{\ast}$}\\
Liu et al. (A) && 234 &52 &142 &34 &$0.10\pm0.03$ &  \multirow{2}{*}{$0.103\pm 0.037^{\ast\ast}$}\\
Shun et al. (A) &&54 &19 &118 &20 &$1.10\pm0.34$ &\\
\hline
\end{tabular}
\caption{\baselineskip=0.5cm{
{\bf Data sets and results of hypothesis testing.} Column 1: Studies selected for the meta--analysis. The index (A) indicates that a possible association was measured in the original publication; the index (NA) indicates that no association was measured in the original publication. Column 2: The phenotype of the patients in the studies grouped by ethnicity. Columns 3--6: Genotypic frequencies of TNF$\alpha$--308 in CHD patients and non--CHD (control) patients from twenty studies (indexed $i$) and for two CHD phenotypes (indexed $j$), namely coronary stenosis (CS) and myocardial infarction (MI). Columns 7--8: The Bayes factors for the hypotheses considered, for each data set $(H^{i,j}_{1}/H^{i,j}_{0}),$ and for the meta--data set of each CHD phenotype $(H^{j}_{1}/H^{j}_{0})$. 
$^{\dagger}$~French cohort. $^{\ddagger}$~Irish cohort. 
$^{\ast}$~Excluding Elahi et al., Dedoussis et al. and Chen et al., respectively for each phenotype.
$^{\ast\ast}$~Excluding Georges et al., Bennet et al. and Hou et al., respectively for each phenotype.}
}
\label{table:ena}
\end{table}

This analysis is based on twenty data sets (indexed $i$) on two CHD phenotypes (indexed $j$) selected from the studies compiled in Ref.~\cite{zhang_2011}, following a well--documented study identification, data acquisition and selection strategy, including also statistical tests (Hardy--Weinberg equilibrium, heterogeneity, publication bias).
The selected data sets are the studies that report the genotypes of both CHD patients and non--CHD (control) patients for the two CHD phenotypes separately. In particular, 
there were included: fifteen data sets from studies on Caucasians, where six studies are on the CHD phenotype coronary stenosis (CS)
\cite{allen_2001,elahi_2008,georges_2003,sbarsi_2008,szalai_2002,vendrell_2003} 
and nine studies are on the CHD phenotype myocardial infarction (MI) \cite{antonicelli_2005,bennet_2006,dedoussis_2005,herrmann_1998,koch_2001,padovani_2000,tobin_2004,tulyakova_2004}; 
and five data sets from studies on Asians on the CHD phenotype coronary stenosis \cite{hou_2009,liu_2009,shun_2009,li_2003, chen_2001}.
The rejected data sets are: three studies on Caucasians (for not reporting data on non--CHD patients); four studies on Asians (three for not reporting data on non--CHD patients and one for not separating the CHD phenotypes); the study on Indians and the study on Africans (both for not separating the CHD phenotypes). 

The data consist of frequencies of occurrence of the --308 TNF--$\alpha$ SNP in randomly selected CHD patients and non--CHD (control) patients, respectively $n_{\text{SNP,CHD}}$ and $n_{\text{SNP},\overline{\text{CHD}}}.$ The data are summarized in Table~\ref{table:ena} (columns 3--6).
The errors indicated were computed from error propagation. Assuming that the methods for measuring the presence of the SNP have a success rate of $r_{\rm{suc}}=0.88$ \cite{pcr_error}, and furthermore that the error of a counting result is given by the Poisson approximation $\sqrt{n},$ then the error of a counting result $n$ on the presence of the SNP is given by $(1-r_{\rm{suc}})\sqrt{n}/2.$ 

\begin{figure}[t] 
\vspace{-0.5cm}
\centerline{
\includegraphics[width=15cm]{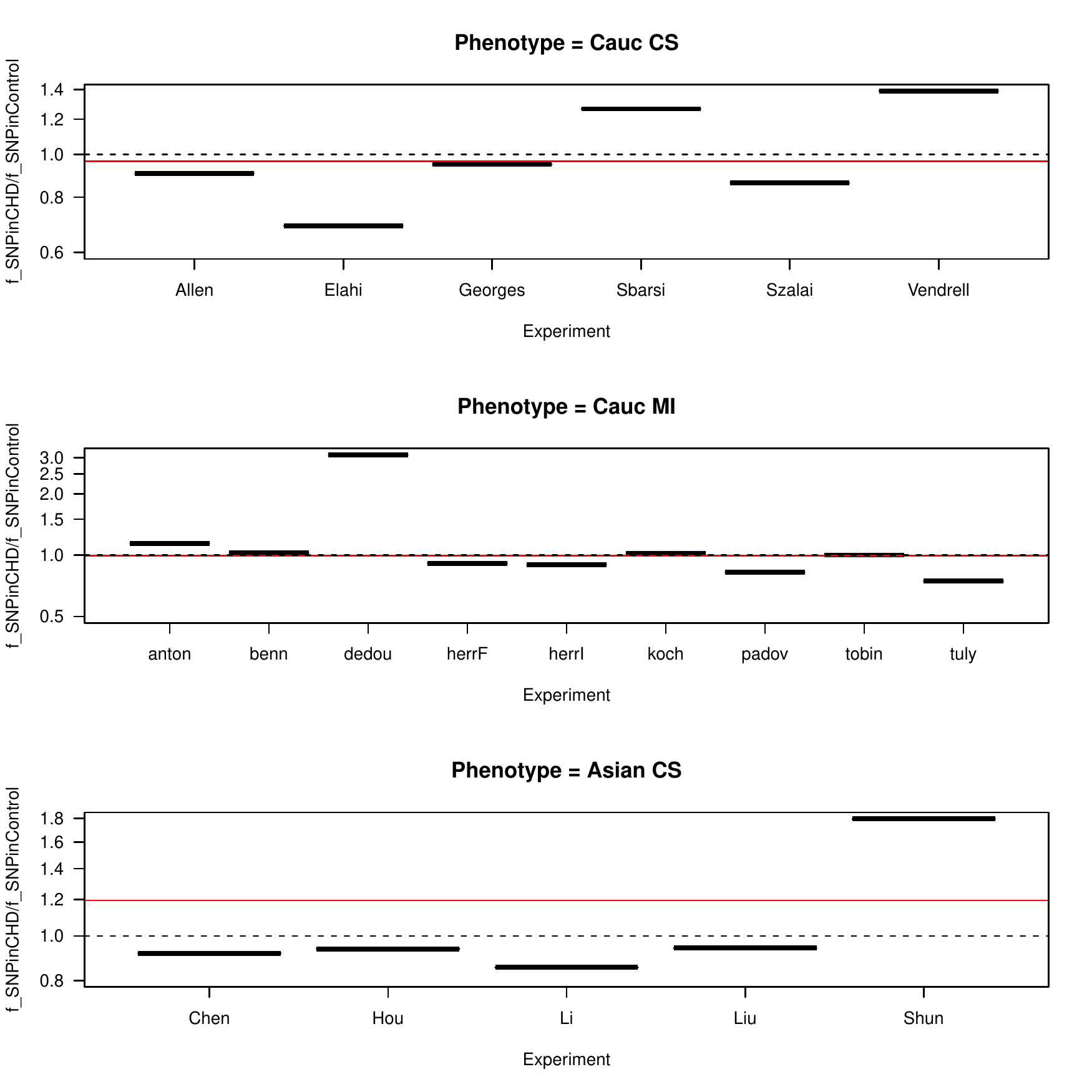} 
}
\vspace{-0.5cm}
\caption{\baselineskip=0.5cm{
{\bf Funnel plot for the ratio of SNP fractions.}  The ratio of the fraction of SNP in the population of CHD patients to the fraction of SNP in the population of non--CHD patients, $f_{\text{SNPinCHD}}/f_{\text{SNPin}\overline{\text{CHD}}}$, for each study, grouped by ethnicity and CHD phenotype. Top panel: Caucasians with coronary stenosis; Middle panel: Caucasians with myocardial infarction; Bottom panel: Asians with coronary stenosis. The solid horizontal line is the ratio of the combined data sets included in each panel. The dashed horizontal line marks the ratio equal to one.}
}
\label{fig:funnel_exp}
\end{figure}

\begin{figure}[t]
\vskip-0.5cm
\centerline{
\includegraphics[width=10cm]{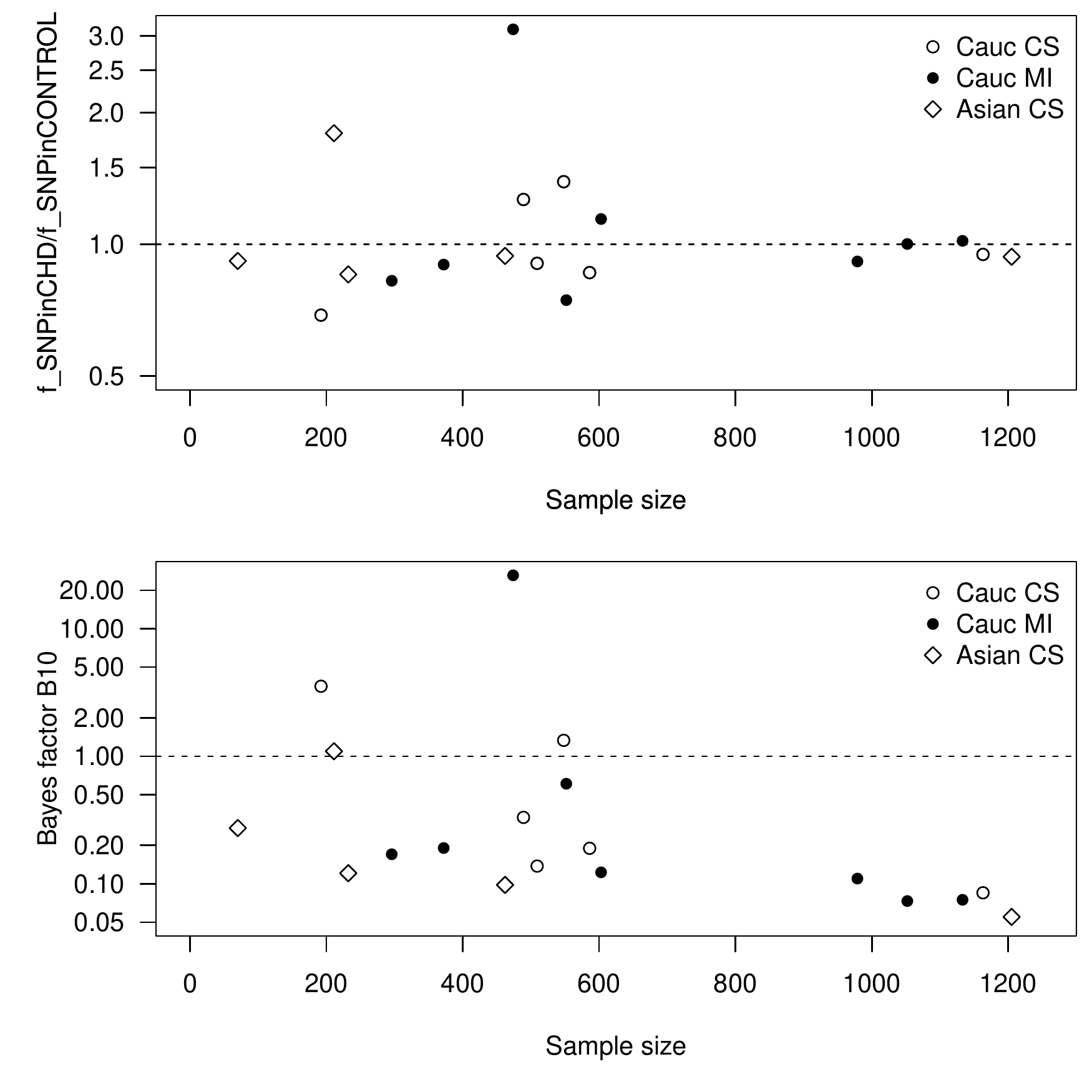}
}
\vskip-0.5cm
\caption{\baselineskip=0.5cm{
{\bf Scatter plots as a function of the sample size.}  Top panel: The ratio of the frequency of SNP in the CHD population to the frequency of SNP in the non--CHD population as a function of the sample size. Bottom panel: The Bayes factor for the two hypotheses discussed in the text as a function of the sample size.}
}
\label{fig:ntotal}
\end{figure}

\subsubsection{Data heterogeneity}

In order to investigate the heterogeneity in the data sets, we compare the size of the effect (defined as a measure of the difference between CHD and non--CHD patients) in each study \cite{walker_2008}. As a measure of the size of the effect, we use the fraction of SNP in the population of CHD patients and in the population of non--CHD patients, respectively $f_{\text{SNPinCHD}}=n_{\text{SNP},\text{CHD}}/n_{\text{CHD}}$ and $f_{\text{SNPin}\overline{\text{CHD}}}=n_{\text{SNP},\overline{\text{CHD}}}/n_{\overline{\text{CHD}}},$ where $n_{\text{CHD}}=n_{\text{SNP},\text{CHD}}+n_{\overline{\text{SNP}},CHD}$ is the total number of CHD patients and  $n_{\overline{\text{CHD}}}=n_{SNP,\overline{\text{CHD}}}+n_{\overline{\text{SNP}},\overline{\text{CHD}}}$ is the total number of non--CHD patients. 
Moreover, the ratio of these two fractions gives an indication of the correlation sign. Hence, if $f_{\text{SNPinCHD}}/f_{\text{SNPin}\overline{\text{CHD}}}>1,$ the SNP is proportionally more frequent in CHD than in non--CHD patients, hence the study favours a positive correlation between the presence of the SNP and the occurrence of CHD; if $f_{\text{SNPinCHD}}/f_{\text{SNPin}\overline{\text{CHD}}}<1,$ the SNP is proportionally less frequent in CHD than in non--CHD patients, hence the study favours a negative correlation; if $f_{\text{SNPinCHD}}/f_{\text{SNPin}\overline{\text{CHD}}}=1,$  the SNP is equally frequent in CHD and in non--CHD patients, hence the study favours no correlation. 

We plot this ratio of fractions for each study, grouped by ethnicity and CHD phenotype, in Fig.~\ref{fig:funnel_exp}. 
We also plot the ratio for the combined data sets included in each panel. We observe that the ratio of the data sets are asymmetrical distributed about the ratio equal to one, showing a predominance of ratios smaller than one. The ratio of the combined data sets included in each panel is slightly smaller than one for the Caucasian studies (for both CHD phenotypes) and larger than one for the Asian studies. This asymmetry indicates heterogeneity in the studies, as also observed in the meta--analysis of Ref.~\cite{zhang_2011}.

In  Fig.~\ref{fig:ntotal} (top panel), we plot this ratio of fractions as a function of the sample size. We observe that smaller data sets are distributed across a wide range of values of this ratio, whereas larger data sets are distributed more closely to one. This implies that smaller data sets favour either positive or negative correlation, whereas larger data sets favour no correlation.

\subsection{Hypotheses Testing}
\label{sec:hypo}

First we test the hypothesis $H_{1}$ that the presence of TNF--$\alpha$ SNP is related to the occurrence of CHD against the null hypothesis $H_{0}$ that the presence of the SNP is unrelated to the occurrence of CHD. 
By the Bayes theorem, the probability of a hypothesis $H_{i}$ given the data $D_{\text{SNP}}$ is the posterior probability of the corresponding hypothesis  
\ba
P(H_{i}\vert D_{\text{SNP}})= { {P(D_{\text{SNP}}\vert H_{i})P(H_{i})}\over P(D_{\text{SNP}})},
\label{eqn:prob_hn}
\ea
where $P(D_{\text{SNP}}\vert H_{i})$ is the evidence, $P(H_{i})$ is the prior probability of $H_{i}$ and $P(D_{\text{SNP}})=\sum_{n}P(D_{\text{SNP}}\vert H_{i})P(H_{i}).$
The subscript in $D_{\text{SNP}}$ reminds us that the random variable is the occurrence of the SNP.
In order to infer which hypothesis is more likely in view of the data, we compare the evidence computed for the two hypotheses. The evidence is the integral of the likelihood over the $j$--dimensional parameter space $p_{i,j}$ of the hypothesis $H_{i}$
\ba
P(D_{\text{SNP}}\vert H_{i})
=\int d^{j}p_{i,j}~P(D_{\text{SNP}}\vert p_{i,j}, H_{i})P(p_{k,n}\vert H_{i}).
\ea
Assuming equal prior probabilities for the two hypotheses, then from Eqn.~(\ref{eqn:prob_hn}) it follows that
\ba
{P(H_{1}\vert D_{\text{SNP}})\over P(H_{0}\vert D_{\text{SNP}})}
={P(D_{\text{SNP}}\vert H_{1})\over P(D_{\text{SNP}}\vert H_{0})}.
\ea
We compute the evidence of the two hypotheses, for each data set separately and for the combined data sets grouped by CHD phenotype. We follow the procedure detailed in Ref~\cite{vourvouhaki_2009}, which we here summarize for one data set and then generalize for the combined data sets. In all cases, we choose a uniform distribution for the prior of the parameters, which is justified by the absence of an {\it a priori} bias on the values of the parameters \cite{mackay}.

The evidence of $H_{0},$ $P(D_{\text{SNP}}\vert H_{0}),$ is computed assuming that the presence of the SNP is described by a binomial distribution with one parameter only, namely the probability $p_0$ that the SNP occurs in a given population. For $n_{\text{SNP}}$ occurrences of the SNP and $n_{\overline{\text{\text{SNP}}}}$ non--occurrences of the SNP in a sample of size $n=n_{\text{SNP}}+n_{\overline{\text{\text{SNP}}}},$ the likelihood $P(D_{\text{SNP}}\vert p_{0}, H_{0})$ is given by
\ba
P(D_{\text{SNP}}\vert p_{0}, H_{0})=p_{0}^{n_{\text{SNP}}}(1-p_{0})^{n_{\overline {\text{SNP}}}}.
\label{eqn:likelihood_h0}
\ea
Moreover, assuming a uniform prior distribution for $p_{0},$ $P(p_{0})=1,$ we find that 
\ba
P(D_{\text{SNP}}\vert H_{0})
=\int _{0}^{1} dp_{0}~P(D_{\text{SNP}}\vert p_{0}, H_{0})P(p_{0}\vert H_{0})
={n_{\text{SNP}}!~n_{\overline{\text{\text{SNP}}}}!\over (n_{\text{SNP}}+n_{\overline{\text{\text{SNP}}}}+1)!},
\label{eqn:h0}
\ea
where $n!$ stands for the factorial of $n.$

The evidence of $H_{1},$ $P(D_{\text{SNP}}\vert H_{1}),$ is computed assuming that the presence of the SNP is described by a binomial distribution with two parameters, namely the probability $p_{1,CHD}$ that the SNP occurs in the subset of CHD  patients and the probability $p_{1,\overline{\text{CHD}}}$ that the SNP occurs in the subset of non--CHD patients,
\ba
P(D_{\text{SNP}}\vert H_{1})
&=&\int _{0}^{1}dp_{1,CHD}\int _{0}^{1}dp_{1,\overline{\text{CHD}}}\cr
&\times&P(D_{\text{SNP}}\vert p_{1,CHD},p_{1,\overline{\text{CHD}}}, H_{1})
P(p_{1,CHD},p_{1,\overline{\text{CHD}}}\vert H_{1}).
\label{eqn:likelihood_h1}
\ea
For  $n_{\text{SNP,CHD}}$ occurrences of the SNP and $n_{\overline{\text{SNP}},\text{CHD}}$ non--occurrences of the SNP in a subset of CHD patients $n_{\text{CHD}}=n_{\text{SNP,CHD}}+n_{\overline{\text{SNP}},\text{CHD}}$, and also for $n_{\text{SNP},\overline{\text{CHD}}}$ occurrences of the SNP and $n_{\overline{\text{SNP}},\overline{\text{CHD}}}$ non--occurrences of the SNP in a subset of non--CHD patients $n_{\overline{\text{CHD}}}=n_{SNP,\overline{\text{CHD}}}+n_{\overline{\text{\text{SNP}}},\overline{\text{CHD}}},$ 
the likelihood $P(D_{\text{SNP}}\vert p_{1,CHD},p_{1,\overline{\text{CHD}}}, H_{1})$ is separable, i.e. it can be decomposed into the product of the likelihoods $P(D_{\text{SNP}}\vert p_{1,CHD}, H_{1})$ and $P(D_{\text{SNP}}\vert p_{1,\overline{\text{CHD}}}, H_{1}),$  as follows
\ba
P(D_{\text{SNP}}\vert p_{1,\text{CHD}},p_{1,\overline{\text{CHD}}}, H_{1})
&=&p_{1,\text{CHD}}^{n_{\text{SNP,CHD}}}(1-p_{1,\text{CHD}})^{n_{\overline {\text{SNP}},\text{CHD}}}\cr
&\times& p_{1,\overline{\text{CHD}}}^{n_{\text{SNP},\overline{\text{CHD}}}}
(1-p_{1,\overline{\text{CHD}}})^{n_{\overline{\text{SNP}},\overline{\text{CHD}}}}\cr
&\equiv& 
P(D_{\text{SNP}}\vert p_{1,\text{CHD}}, H_{1})
P(D_{\text{SNP}}\vert p_{1,\overline{\text{CHD}}}, H_{1}).
\ea
Assuming a uniform probability for $p_{1,\text{CHD}}$ and $p_{1,\overline{\text{CHD}}},$ 
$P(p_{1,\text{CHD}},p_{1,\overline{\text{CHD}}}\vert H_{1})=$1 
and moreover that the priors on $p_{1,\text{CHD}}$ and $p_{1,\overline{\text{CHD}}}$  are separable, 
the posterior distribution will also be separable and given by
\ba
P(D_{\text{SNP}}\vert H_{1})
&=&
\int _{0}^{1} dp_{1,\text{CHD}}~P(D_{\text{SNP}}\vert p_{1,\text{CHD}}, H_{1})P( p_{1,\text{CHD}}\vert H_{1})\cr
&\times&\int _{0}^{1} dp_{1,\overline{\text{CHD}}}~
P(D_{\text{SNP}}\vert p_{1,\overline{\text{CHD}}}, H_{1})P(p_{1,\overline{\text{CHD}}}\vert H_{1})\cr
&=&{n_{\text{SNP,CHD}}!~n_{\overline{\text{SNP}},\text{CHD}}!\over (n_{\text{SNP,CHD}}+n_{\overline{\text{SNP}},\text{CHD}}+1)!}
{n_{\text{SNP},\overline{\text{CHD}}}!~n_{\overline{\text{SNP}},\overline{\text{CHD}}}!\over
(n_{\text{SNP},\overline{\text{CHD}}}+n_{\overline{\text{SNP}},\overline{\text{CHD}}}+1)!}.
\label{eqn:h1}
\ea

In order to compare the hypotheses, we take the ratio of the corresponding evidences, $B_{10}=P(H_{1}\vert D)/P(H_{0}\vert D),$ which we present in Table~\ref{table:ena} (columns 7--8). This quantity is known as the Bayes factor and gives empirical levels of significance for the strength of the evidence of the test hypothesis over that of the null hypothesis. It also encapsulates the Occam's factor, which measures the adequacy of a hypothesis to the data over the parameter space of the hypothesis \cite{mackay}. 
The levels of significance ascribed to the Bayes factor are calibrated by the Jeffrey's scale \cite{jeffrey}. According to this scale, a Bayes factor larger than one indicates that $H_{1}$ is favoured over $H_{0}.$ Otherwise, $H_{0}$ is favoured over $H_{1}.$ For the data sets taken separately, the results from this hypothesis test mostly agree with the corresponding results presented in the meta--analysis by Chu et al. (see Fig.~1 of Ref.~\cite{chu_2013}).

\begin{figure}[t] 
\vspace{-0.5cm}
\centerline{
\includegraphics[width=15cm]{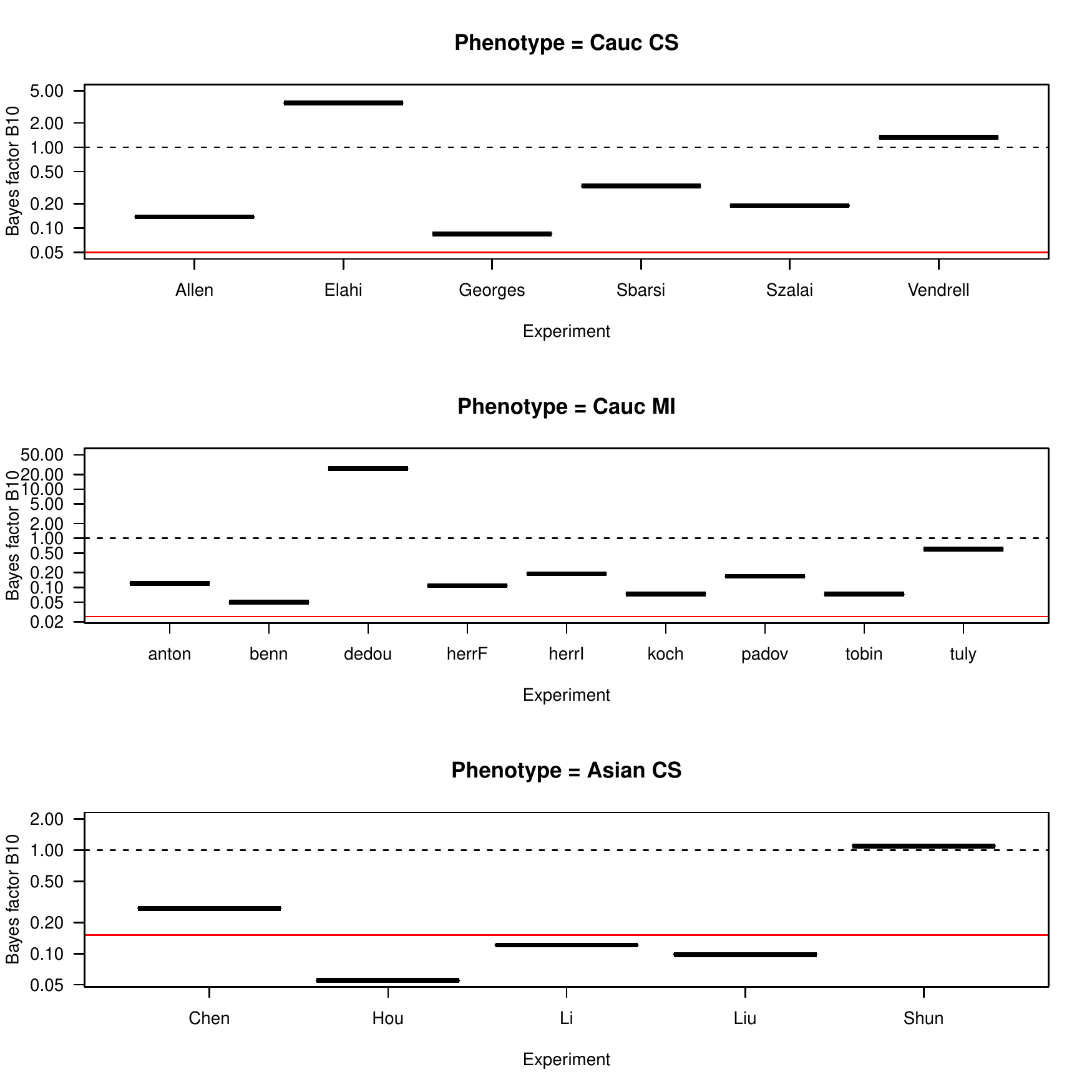} 
}
\vspace{-0.5cm}
\caption{\baselineskip=0.5cm{
{\bf Funnel plot for the Bayes factor.} The Bayes factor for each study, grouped by ethnicity and CHD phenotype. Top panel: Caucasians with coronary stenosis; Middle panel: Caucasians with myocardial infarction; Bottom panel: Asians with coronary stenosis. The solid horizontal line marks the average Bayes factor of the data sets included in each panel. The dashed horizontal line marks the Bayes factor equal to one.}
}
\label{fig:bayes_exp}
\end{figure}

We plot the Bayes factor for each study, grouped by ethnicity and CHD phenotype, in Fig.~\ref{fig:bayes_exp}.  For the data sets taken separately, we observe that the Bayes factor is asymmetrically distributed about the Bayes factor equal to one, with most Bayes factors being smaller than one. 
The exceptions are Elahi et al. \cite{elahi_2008}, Vendrell et al. \cite{vendrell_2003} and Dedoussis et al. \cite{dedoussis_2005} for the Caucasian population, and Shun et al. \cite{shun_2009} for the Asian population. 
This asymmetry 
indicates heterogeneity in the results.
For the combined data sets included in each panel, the Bayes factor takes values $0.03-0.05$ for the Caucasian population and $0.15$ for the Asian population, which indicates that there is no evidence for $H_{1}$ over $H_{0}.$ 
We also observe that, for the Caucasian population, the Bayes factor of the combined data sets is outside the range of variability of the Bayes factor of the data sets considered separately. This suggests that the combination of the Caucasian data sets causes a new data pattern to emerge. 
Conversely the combination of the Asian data sets leads to an approximately average data pattern.
Hence we conclude that the data favour $H_{0}$ over $H_{1}.$ Since $H_{0}$ yields trivial results, in the subsequent subsections we present the results also for $H_{1}$ to illustrate the application of the formalism to a more general setup. It is also instructive to compare the subsequent results using both hypotheses.

In Fig.~\ref{fig:ntotal} (bottom panel), we plot the Bayes factor as a function of the sample size. We observe that smaller data sets are distributed across a wide range of values of the Bayes factor, whereas larger data sets are distributed across values smaller than one. This implies that smaller data sets favour either $H_{0}$ or $H_{1},$ whereas larger data sets favour $H_{0}.$ 

\begin{figure}[t] 
\vskip-0.5cm
\centerline{
\includegraphics[width=10cm]{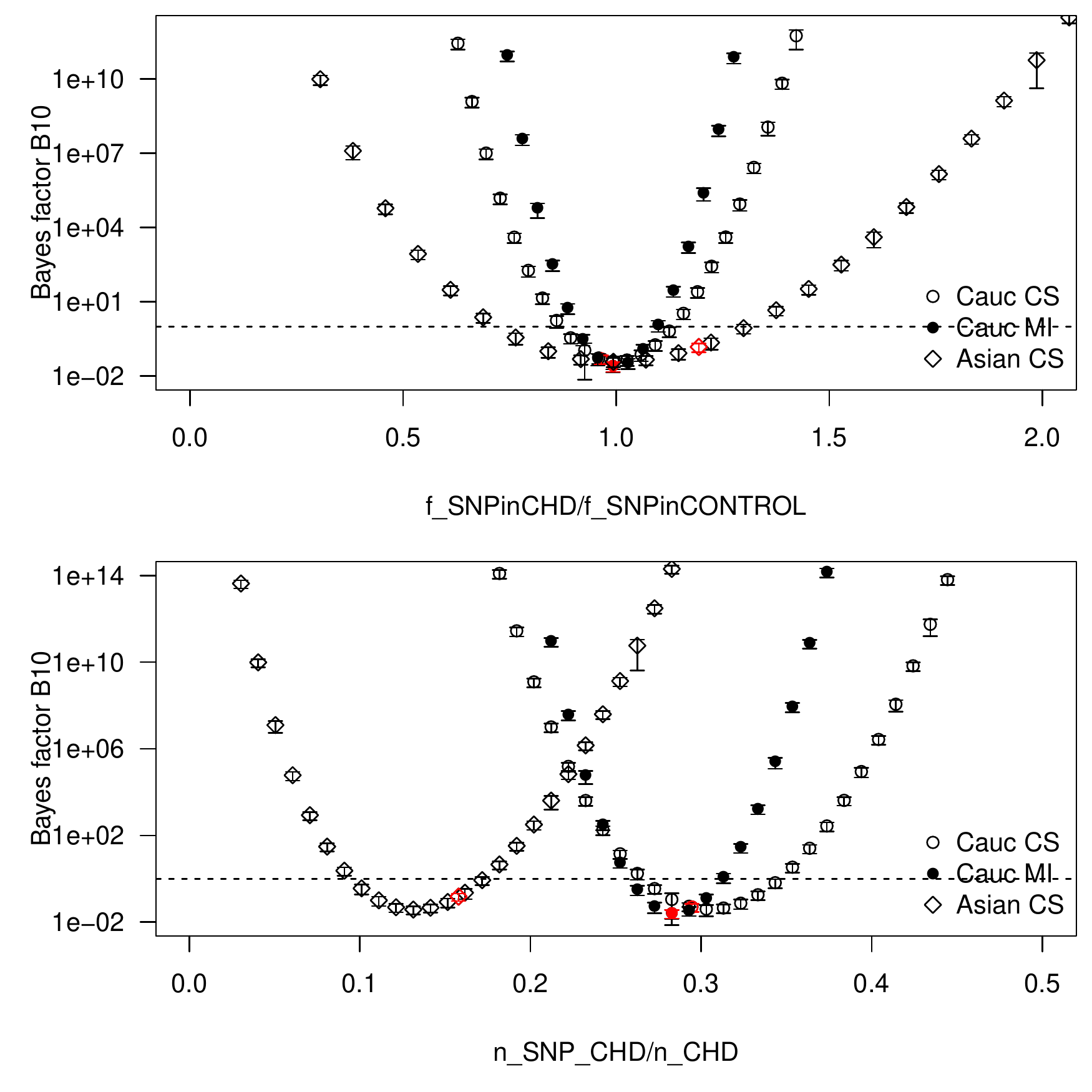}
}
\caption{\baselineskip=0.5cm{
{\bf Bayes factor as a function of the frequency of SNP in the CHD populations.} The Bayes factor for several realizations of CHD populations with the same $n_{\text{CHD}}$ but with different fractions of SNP, grouped by ethnicity and CHD phenotype. The realizations that correspond to a real combined data set are marked as red points. The dashed horizontal line marks the Bayes factor equal to one. Top panel: The Bayes factor as a function of $f_{\text{SNPinCHD}}/f_{\text{SNPin}\overline{\text{CHD}}}.$ Bottom panel: The Bayes factor as a function of $f_{\text{SNPinCHD}}.$}
}
\label{fig:bayes_arr}
\end{figure}

\subsubsection{Correlation sign}
Comparing Fig.~\ref{fig:bayes_exp} with Fig.~\ref{fig:funnel_exp}, we observe that, among the studies with Bayes factor larger than one, Elahi et al. has a ratio $f_{\text{SNPinCHD}}/f_{\text{SNPin}\overline{\text{CHD}}}<1,$ i.e. the SNP is proportionally less frequent in CHD than in non--CHD patients, which indicates a negative correlation between the presence SNP and the occurrence of CHD. Another example of comparatively large Bayes factor and low ratio $f_{\text{SNPinCHD}}/f_{\text{SNPin}\overline{\text{CHD}}}$ is the study of Tuliakova et al. This indicates that the hypotheses as formulated do not distinguish the correlation sign.  

To further explore how the ratio $f_{\text{SNPinCHD}}/f_{\text{SNPin}\overline{\text{CHD}}}$ 
affects the result of the hypothesis testing, we consider several realizations of CHD populations with the same $n_{\text{CHD}}$  but with different fractions of SNP. More specifically for each combined data set, we vary $n_{\text{SNP,CHD}}$ while varying simultaneously $n_{\overline{\text{SNP}},\text{CHD}}$ so as to keep $n_{\text{CHD}}$ constant. Throughout the different realizations, the size of the control population is kept equal to the size of the control population of the combined data sets grouped by ethnicity and CHD phenotype. For each realization, we compute both $f_{\text{SNPinCHD}}$ (note that $f_{\text{SNPin}\overline{\text{CHD}}}$ is by construction kept fixed) and $B_{10},$ and plot the results in Fig.~\ref{fig:bayes_arr}. The realizations with the $f_{\text{SNPinCHD}}$ of a real combined data set are marked as red points. In the top panel, we plot $B_{10}$ as a function of $f_{\text{SNPinCHD}}/f_{\text{SNPin}\overline{\text{CHD}}},$ from which there result three parabolae centred at the same point. 
In the bottom panel, for a better visualization of the behaviour of $B_{10},$ we plot $B_{10}$ as a function of $f_{\text{SNPinCHD}},$ from which there result three parabolae centred at different points. 
We observe that $B_{10}$ follows a parabola, taking the minimum value when $f_{\text{SNPinCHD}}/f_{\text{SNPin}\overline{\text{CHD}}}=1$ and increasing in both directions with the increase of $\vert f_{\text{SNPinCHD}}/f_{\text{SNPin}\overline{\text{CHD}}}-1\vert,$ i.e. with the increase of the distance from 1. This confirms that the hypotheses as formulated do not distinguish between a positive correlation of the SNP with CHD ($f_{\text{SNPinCHD}}/f_{\text{SNPin}\overline{\text{CHD}}}>1$) and a negative correlation ($f_{\text{SNPinCHD}}/f_{\text{SNPin}\overline{\text{CHD}}}<1$). Hence, the value of $f_{\text{SNPinCHD}}/f_{\text{SNPin}\overline{\text{CHD}}}$ complements the value of $B_{10}$ in the characterization of the correlation. 

\section{Results}
\label{sec:results}

\subsection{Inference of conditional probabilities}
\label{sec:infer}

\subsubsection{Posterior probability for the occurrence of CHD}
\label{sec:post}

We proceed to compute the probability for the occurrence of (\text{CHD,} i.e. given the data on the presence of the SNP, we determine the probability that a patient has CHD. This is defined as the posterior probability 
\ba
P(\text{CHD}\vert D_{\text{SNP}}, H_{i})
={P(D_{\text{SNP}}\vert (\text{CHD,} H_{i})P(\text{CHD})\over P(D_{\text{SNP}}\vert H_{i})}.
\label{eqn:p_chd_post}
\ea 

The prior probability $P(\text{CHD})$ is based on the available information on the occurrence of CHD. This probability can be computed by combining all the risk factors per age interval per pathology. According to the European guidelines, less than 4 in 1000 people have CS \cite{esc_cs}, whereas about 1 in 1000 people have MI \cite{esc_mi}. 
We then use $P(\text{CHD})=0.004$ for CS and $P(\text{CHD})=0.001$ for MI. 

The evidence $P(D_{\text{SNP}}\vert H_{i})$ can be decomposed as
\ba
P(D_{\text{SNP}}\vert H_{i})=P(D_{\text{SNP}}\vert (\text{CHD,} H_{i}) P(\text{CHD})
+P(D_{\text{SNP}}\vert \overline{\text{CHD}}, H_{i})P(\overline{\text{CHD}}).
\ea
In the case of $H_{0},$ 
\ba
P(D_{\text{SNP}}\vert (\text{CHD,} H_{0})
&=&{{n}\choose n_{\text{SNP}}}
p_{0}^{n_{\text{SNP}}}(1-p_{0})^{n_{\overline{\text{SNP}}}} \equiv P(D_{\text{SNP}}\vert H_{0})\cr
P(D_{\text{SNP}}\vert  \overline{\text{CHD}}, H_{0})&=&P(D_{\text{SNP}}\vert H_{0}),
\label{eqn:evidence_h0}
\ea
whereas in the case of $H_{1},$
\ba
P(D_{\text{SNP}}\vert (\text{CHD,} H_{1})
&=&{n_{\text{CHD}}\choose n_{\text{SNP,CHD}}}
p_{1,\text{CHD}}^{n_{\text{SNP,CHD}}}(1-p_{1,CHD})^{n_{\overline{\text{SNP}},\text{CHD}}}\cr
P(D_{\text{SNP}}\vert \overline{\text{CHD}}, H_{1})
&=&{n_{\overline{\text{CHD}}}\choose n_{SNP,\overline{\text{CHD}}}} 
p_{1,\overline{\text{CHD}}}^{n_{\text{SNP},\overline{\text{CHD}}}}(1-p_{1,\overline{\text{CHD}}})^{n_{\overline{\text{SNP}},\overline{\text{CHD}}}}.
\label{eqn:evidence_h1}
\ea
In the previous section, we computed the evidence by marginalizing the parameters of each hypothesis. Here, assuming a hypothesis $H_{i}$ and 
using the Bayes theorem, we compute the posterior probability of each parameter $p_{i,j}$ given the data 
\ba
P(p_{i,j}\vert D_{\text{SNP}})={P(D_{\text{SNP}}\vert p_{i,j})P(p_{i,j}\vert H_{i})\over P(D_{\text{SNP}}\vert H_{i})},
\ea
and find for  $p_{i,j}$  the value that maximizes the likelihood $P(D_{\text{SNP}}\vert p_{i,j}).$
In the case of $H_{0},$ we compute the posterior probability of the single parameter $p_{i,j}=p_{0},$ where
$P(D_{\text{SNP}}\vert p_{0})$ is given by Eqn.~(\ref{eqn:likelihood_h0}), $P(D_{\text{SNP}}\vert H_{0})$ is given by Eqn.~(\ref{eqn:h0}) and $P(p_{0}\vert H_{0})$ is assumed uniform. Taking the derivative with respect to $p_{0}$ and solving for $dP(p_{0}\vert D_{\text{SNP}})/dp_{0}=0,$ we find for the maximum--likelihood value of $p_{0}$ the value 
\ba 
p_{0 (\text{maxL})}={n_{\text{SNP}}\over (n_{\text{SNP}}+n_{\overline{\text{\text{SNP}}}})}.
\ea
Similarly in the case of $H_{1},$ we compute the posterior probability of each of the two parameters $p_{i,j}=\{p_{1,\text{CHD}}, p_{1,\overline{\text{CHD}}}\},$ where $P(D_{\text{SNP}}\vert p_{1,\text{CHD}})$ and  $P(D_{\text{SNP}}\vert p_{1,\overline{\text{CHD}}})$ are given by Eqn.~(\ref{eqn:likelihood_h1}), $P(D_{\text{SNP}}\vert H_{1})$ is given by Eqn.~(\ref{eqn:h1}) and both $P(p_{1,\text{CHD}}\vert H_{1})$ and $P(p_{1,\overline{\text{CHD}}}\vert H_{1})$ are assumed uniform, finding for the maximum--likelihood values of $p_{1,\text{CHD}}$ and $p_{1,\overline{\text{CHD}}}$ respectively
\ba 
p_{1,\text{CHD}(\text{maxL})}&=&{n_{\text{SNP,CHD}}\over (n_{\text{SNP,CHD}}+n_{\overline{\text{SNP}},\text{CHD}})}, \\
p_{1,\overline{\text{CHD}} (\text{maxL})}&=&{n_{\text{SNP},\overline{\text{CHD}}}\over 
  (n_{\text{SNP},\overline{\text{CHD}}}+n_{\overline{\text{SNP}},\overline{\text{CHD}}})}.
\ea

Analogously we define the posterior probability
\ba
P(\overline{\text{CHD}}\vert D_{\text{SNP}}, H_{i})
={P(D_{\text{SNP}}\vert \overline{\text{CHD}}, H_{i})P(\overline{\text{CHD}})\over P(D_{\text{SNP}}\vert H_{i})}.
\label{eqn:p_chd_post}
\ea 

Finally, using the maximum--likelihood value of $p_{i,j},$ we compute $P(\text{CHD}\vert D_{\text{SNP}}, H_{i})$ for the data sets 
combined, which we present in Table~\ref{table:dyo}.

In the case of $H_{0},$ no information is added to the posterior probability, since by Eqn.~(\ref{eqn:evidence_h0}) the posterior probabilities equal the prior. Conversely in the case of $H_{1},$ information is added to the posterior probability, since by Eqn.~(\ref{eqn:evidence_h1}) there result posterior probabilities different from the prior albeit compatible with the prior.

\begin{table}[t]
\begin{tabular}{c|c|ccc}
Hypothesis 
& Probabilities
&\multicolumn{3}{c}{Phenotype ($j$)}\\ \cline{3-5}
&& Cauc CS & Cauc MI & Asian CS\\ 
 \hline\hline
\multirow{5}{*}{$H_{0}$} 
& $p_{0}$ & $0.299\pm0.001$ & $0.284\pm0.001$ & $0.141\pm0.001$ \\
&$P(\text{CHD}\vert D_{\text{SNP}}, H_{0})$ 
  & $(4.00\pm 1.31)\cdot 10^{-3}$ & $(1.00\pm 0.25)\cdot 10^{-3}$ & $(4.00\pm 0.91)\cdot 10^{-3}$\\
&$P(\text{nextSNP}, CHD \vert D_{\text{SNP}}, H_{0})$ 
  & $(1.19\pm0.39)\cdot10^{-3}$ & $(0.28\pm0.07)\cdot10^{-3}$ & $(0.56\pm 0.13)\cdot 10^{-3}$\\
&$P(\text{nextSNP}, \overline{\text{CHD}} \vert D_{\text{SNP}}, H_{0})$ 
  & $0.298\pm1.093$ & $0.284\pm1.752$ & $0.141\pm0.360$\\
&$P(\text{nextSNP}\vert D_{\text{SNP}}, H_{0})$  
  & $0.299\pm1.093$  & $0.284\pm1.572$ & $0.141\pm0.360$\\
& $r_{\text{nextSNP,CHD}}$
& $(4.00\pm14.65)\cdot 10^{-3}$  & $(1.00\pm5.54)\cdot 10^{-3}$ & $(4.00\pm10.22)\cdot 10^{-3}$\\
 \hline
 \multirow{6}{*}{$H_{1}$} 
& $p_{1,CHD}$ & $0.295\pm0.001$ & $0.283\pm0.001$ & $0.158\pm 0.001$\\
& $p_{1,\overline{\text{CHD}}}$ & $0.305\pm0.001$ & $0.285\pm0.001$ & $0.132\pm 0.001$\\
& $P(\text{CHD}\vert D_{\text{SNP}},H_{1})$ 
  & $(3.42\pm7.94)\cdot 10^{-3}$ & $(0.98\pm3.26)\cdot 10^{-3}$ & $(5.00\pm 7.02)\cdot 10^{-3}$\\
& $P(\text{nextSNP}, CHD \vert D_{\text{SNP}}, H_{1})$ 
  & $(1.00\pm2.34)\cdot 10^{-3}$ & $(0.28\pm0.92)\cdot 10^{-3}$ & $(0.79\pm1.11)\cdot 10^{-3}$\\
& $P(\text{nextSNP}, \overline{\text{CHD}} \vert D_{\text{SNP}}, H_{1})$ 
  & $0.304\pm0.598$ & $0.285\pm0.926$ & $0.131\pm0.244$\\
& $P(\text{nextSNP}\vert D_{\text{SNP}}, H_{1})$  
  & $0.305\pm0.598$ & $0.285\pm0.926$ & $0.132\pm 0.244$ \\
&$r_{\text{nextSNP,CHD}}$
& $(3.30\pm10.02)\cdot 10^{-3}$ & $(0.98\pm4.54)\cdot 10^{-3}$ & $(6.00\pm 13.84)\cdot 10^{-3}$\\
\hline
\end{tabular}
\caption{\baselineskip=0.5cm{
{\bf Probabilities inferred from the combined data sets.} 
To each hypothesis there correspond several rows consisting of : a) the parameters $p_{i,j}$ given by the maximum--likelihood values, in particular, $p_{0}$ (hence one row) in the case of $H_{0},$ $p_{1,\text{CHD}}$ and $p_{1,\overline{\text{CHD}}}$ (hence two rows) in the case of $H_{1};$ b) the posterior probability for the occurrence of (\text{CHD,} $P(\text{CHD}\vert D_{\text{SNP}}, H_{i})$ (hence one row for each hypothesis);  c) the predicted probabilities for the presence of the SNP, namely $P(\text{nextSNP},\text{CHD} \vert D_{\text{SNP}}, H_{i}),$ $P(\text{nextSNP}, \overline{\text{CHD}} \vert D_{\text{SNP}}, H_{i})$ and $P(\text{nextSNP}\vert D_{\text{SNP}}, H_{i})$ (hence three rows for each hypothesis); and d) the probability ratio that measures the influence of CHD in the presence of the SNP, $r_{\text{nextSNP,CHD}}\equiv P(\text{nextSNP}, \text{CHD} \vert D_{\text{SNP}},H_{i})/P(\text{nextSNP}\vert D_{\text{SNP}}, H_{i}),$ computed from the combined data of each phenotype (hence one row for each hypothesis).  Column 1: The hypotheses. Column 2: The inferred quantities, as described above. Columns 3--5: The values of the inferred quantities for the combined ethnicity and CHD phenotype.}
}
\label{table:dyo} 
\end{table}

\subsubsection{Prediction of the presence of the SNP}
\label{sec:pred}

We now proceed to compute the probability for the presence of the SNP, i.e. given the data, we determine the probability that a randomly selected patient (with or without CHD) has the SNP. This probability is defined as 
\ba
P(\text{nextSNP}\vert D_{\text{SNP}}, H_{i})
&=&P(\text{nextSNP}\vert D_{\text{SNP}},\text{CHD})P(\text{CHD}\vert D_{\text{SNP}}, H_{i})\cr
&+&P(\text{nextSNP}\vert D_{\text{SNP}}, \overline{\text{CHD}})P(\overline{\text{CHD}}\vert D_{\text{SNP}}, H_{i})\cr
&\equiv& P(\text{nextSNP},\text{CHD}\vert D_{\text{SNP}},H_{i})+P(\text{nextSNP}, \overline{\text{CHD}}\vert D_{\text{SNP}},H_{i}).\qquad
\label{eqn:p_next_snp}
\ea
In the case of $H_{0},$ 
\ba
P(\text{nextSNP}\vert D_{\text{SNP}}, \text{CHD})=P(\text{nextSNP}\vert D_{\text{SNP}}, \overline{\text{CHD}})=p_{0},
\label{eqn:p_next_snp_given_chd_h0}
\ea
whereas in the case of $H_{1},$
\ba
P(\text{nextSNP}\vert D_{\text{SNP}}, \text{CHD})=p_{1,\text{CHD}},\cr
P(\text{nextSNP}\vert D_{\text{SNP}}, \overline{\text{CHD}})=p_{1,\overline{\text{CHD}}}.
\label{eqn:p_next_snp_given_chd_h1}
\ea
Using the maximum--likelihood values of $p_{i,j}$ and the posterior probability $P(\text{CHD}\vert D_{\text{SNP}}, H_{i})$ computed above, we compute $P(\text{nextSNP}\vert D_{\text{SNP}}, H_{i}),$ which we present in Table \ref{table:dyo}.

For completion, using the Bayes theorem, we invert $P(\text{nextSNP}\vert D_{\text{SNP}},\text{CHD})$ to find the probability that CHD will occur given that the SNP is present in a randomly selected patient
\ba
P(\text{CHD}\vert \text{nextSNP}, H_{i})={
  P(\text{nextSNP}\vert D_{\text{SNP}},\text{CHD}) P(\text{CHD}\vert D_{\text{SNP}}, H_{i}) 
  \over P(\text{nextSNP}\vert D_{\text{SNP}}, H_{i})
  }.
\label{eqn:p_chd_given_next_snp}
\ea
Similarly, inverting $P(\text{nextSNP}\vert D_{\text{SNP}}, \overline{\text{CHD}}),$ we find the probability that CHD will not occur given that the SNP is present in a randomly selected patient,
$P(\overline{\text{CHD}}\vert \text{nextSNP}, H_{i}),$ which can be found simply by replacing CHD by $\overline{\text{CHD}}$ in 
Eqn.~(\ref{eqn:p_chd_given_next_snp}). 

In order to quantify the influence of CHD in the presence of the SNP, we compute the ratio of 
$P(\text{nextSNP},\text{CHD}\vert D_{\text{SNP}})$  
to $P(\text{nextSNP}\vert D_{\text{SNP}}, H_{i}),$ which gives an estimate of how much the occurrence of CHD indicates the presence of the SNP. This is also the probability in Eqn.~(\ref{eqn:p_chd_given_next_snp}). In the case of $H_{0},$ this ratio equals the posterior probability of occurrence of CHD. Conversely in the case of $H_{1},$ this ratio is different from the posterior probability of occurrence of CHD albeit compatible with it. The occurrence of CHD indicates the presence of the SNP in of order $0.1\%$ of patients ($0.1-0.4\%$ in the case of $H_{0},$ $0.1-0.6\%$ in the case of $H_{1}$), which suggests that the occurrence of CHD is not a good marker for the presence of the SNP. 

In order to quantify the influence of the SNP in the occurrence of (\text{CHD,} we compute the ratio of 
$P(\text{CHD}\vert \text{nextSNP}, H_{i})$ to $P(\text{CHD}\vert D_{\text{SNP}}, H_{i}),$ which gives an estimate of how much the presence of the SNP indicates the occurrence of CHD. This is also the probability in Eqns.~(\ref{eqn:p_next_snp_given_chd_h0}, \ref{eqn:p_next_snp_given_chd_h1}). The presence of SNP indicates the occurrence of CHD in of order $0.1\%$ of patients  ($0.141-0.299.\%$ in the case of $H_{0},$ $0.158-0.295\%$ in the case of $H_{1}$), which suggests that the presence of the SNP is not a risk factor for the emergence of CHD.

\begin{table}[t]
\begin{tabular}{c|c|ccc}
Hypothesis 
& Probabilities
&\multicolumn{3}{c}{Phenotype ($j$)}\\ \cline{3-5}
&& Cauc CS & Cauc MI & Asian CS\\ 
 \hline\hline
\multirow{5}{*}{$H_{0}$} 
& $p_{0}$ & $0.288\pm0.001$ & $0.296\pm0.001$ & $0.136\pm0.001$ \\
&$P(\text{CHD}\vert D_{\text{SNP}}, H_{0})$ 
  & $(4.00\pm 1.26)\cdot 10^{-3}$ & $(1.00\pm 0.24)\cdot 10^{-3}$ & $(4.00\pm 0.89)\cdot 10^{-3}$\\
&$P(\text{nextSNP}, CHD \vert D_{\text{SNP}}, H_{0})$ 
  & $(1.15\pm0.36)\cdot10^{-3}$ & $(0.30\pm0.07)\cdot10^{-3}$ & $(0.55\pm 0.12)\cdot 10^{-3}$\\
&$P(\text{nextSNP}, \overline{\text{CHD}} \vert D_{\text{SNP}}, H_{0})$ 
  & $0.287\pm1.018$ & $0.296\pm1.605$ & $0.136\pm0.340$\\
&$P(\text{nextSNP}\vert D_{\text{SNP}}, H_{0})$  
  & $0.289\pm1.018$  & $0.296\pm1.605$ & $0.136\pm0.340$\\
& $r_{\text{nextSNP,CHD}}$  & $(4.00\pm14.16)\cdot 10^{-3}$  & $(1.00\pm5.54)\cdot 10^{-3}$ & $(4.00\pm10.00)\cdot 10^{-3}$\\
 \hline
 \multirow{6}{*}{$H_{1}$} 
& $p_{1,CHD}$ & $0.290\pm0.001$ & $0.292\pm0.001$ & $0.151\pm 0.001$\\
& $p_{1,\overline{\text{CHD}}}$ & $0.287\pm0.001$ & $0.300\pm0.001$ & $0.128\pm 0.001$\\
& $P(\text{CHD}\vert D_{\text{SNP}},H_{1})$ 
  & $(3.34\pm7.57)\cdot 10^{-3}$ & $(0.99\pm3.18)\cdot 10^{-3}$ & $(5.11\pm 6.96)\cdot 10^{-3}$\\
& $P(\text{nextSNP}, CHD \vert D_{\text{SNP}}, H_{1})$ 
  & $(0.97\pm2.19)\cdot 10^{-3}$ & $(0.29\pm0.93)\cdot 10^{-3}$ & $(0.77\pm1.05)\cdot 10^{-3}$\\
& $P(\text{nextSNP}, \overline{\text{CHD}} \vert D_{\text{SNP}}, H_{1})$ 
  & $0.286\pm0.542$ & $0.300\pm0.947$ & $0.128\pm0.234$\\
& $P(\text{nextSNP}\vert D_{\text{SNP}}, H_{1})$  
  & $0.287\pm0.543$ & $0.300\pm0.947$ & $0.129\pm 0.234$ \\
& $r_{\text{nextSNP,CHD}}$
& $(3.38\pm 9.96)\cdot 10^{-3}$ & $(0.96\pm4.34)\cdot 10^{-3}$ & $(6.02\pm 13.67)\cdot 10^{-3}$\\
\hline
\end{tabular}
\caption{\baselineskip=0.5cm{
{\bf Probabilities inferred from the combined data sets excluding the low--significance data sets and the data sets with extreme results.} Excluded: Elanhi et al. \cite{elahi_2008}, Dedoussis et al. \cite{dedoussis_2005} and Chen et al. \cite{chen_2001}. 
Similarly to Table~\ref{table:dyo}, 
to each hypothesis there correspond several rows consisting of: a) the parameters given by the maximum--likelihood values (one row in the case of $H_{0}$ and two rows in the case of $H_{1}$); b) the posterior probability for the occurrence of CHD (one row for each hypothesis); c) the predicted probabilities for the presence of the SNP (three rows for each hypothesis); and d) the probability ratio that measures the influence of CHD in the presence of the SNP, computed from the combined data of each phenotype  (one row for each hypothesis).  Column 1: The hypotheses. Column 2: The inferred quantities, as described above. Columns 3--5: The values of the inferred quantities for the combined ethnicity and CHD phenotype.}
}
\label{table:dyo_improb} 
\end{table}

\begin{table}[t]
\begin{tabular}{c|c|ccc}
Hypothesis 
& Probabilities
&\multicolumn{3}{c}{Phenotype ($j$)}\\ \cline{3-5}
&& Cauc CS & Cauc MI & Asian CS\\ 
 \hline\hline
\multirow{5}{*}{$H_{0}$} 
& $p_{0}$ & $0.308\pm0.001$ & $0.271\pm0.001$ & $0.177\pm0.001$ \\
&$P(\text{CHD}\vert D_{\text{SNP}}, H_{0})$ 
  & $(4.00\pm 1.08)\cdot 10^{-3}$ & $(1.00\pm 0.20)\cdot 10^{-3}$ & $(4.00\pm 0.63)\cdot 10^{-3}$\\
&$P(\text{nextSNP}, CHD \vert D_{\text{SNP}}, H_{0})$ 
  & $(1.12\pm0.33)\cdot10^{-3}$ & $(0.27\pm0.05)\cdot10^{-3}$ & $(0.71\pm 0.11)\cdot 10^{-3}$\\
&$P(\text{nextSNP}, \overline{\text{CHD}} \vert D_{\text{SNP}}, H_{0})$ 
  & $0.306\pm0.923$ & $0.270\pm1.220$ & $0.177\pm0.314$\\
&$P(\text{nextSNP}\vert D_{\text{SNP}}, H_{0})$  
  & $0.308\pm0.923$  & $0.271\pm1.220$ & $0.177\pm0.314$\\
&$r_{\text{nextSNP,CHD}}$
& $(4.00\pm12.05)\cdot 10^{-3}$  & $(1.00\pm4.51)\cdot 10^{-3}$ & $(4.00\pm7.11)\cdot10^{-3}$\\
 \hline
 \multirow{6}{*}{$H_{1}$} 
& $p_{1,CHD}$ & $0.306\pm0.001$ & $0.270\pm0.001$ & $0.190\pm 0.001$\\
& $p_{1,\overline{\text{CHD}}}$ & $0.309\pm0.001$ & $0.271\pm0.001$ & $0.165\pm 0.001$\\
& $P(\text{CHD}\vert D_{\text{SNP}},H_{1})$ 
  & $(3.93\pm6.97)\cdot 10^{-3}$ & $(0.92\pm2.58)\cdot 10^{-3}$ & $(3.90\pm 4.35)\cdot 10^{-3}$\\
& $P(\text{nextSNP}, CHD \vert D_{\text{SNP}}, H_{1})$ 
  & $(1.20\pm2.14)\cdot 10^{-3}$ & $(0.25\pm0.70)\cdot 10^{-3}$ & $(0.74\pm0.828)\cdot 10^{-3}$\\
& $P(\text{nextSNP}, \overline{\text{CHD}} \vert D_{\text{SNP}}, H_{1})$ 
  & $0.308\pm0.535$ & $0.271\pm0.698$ & $0.165\pm0.187$\\
& $P(\text{nextSNP}\vert D_{\text{SNP}}, H_{1})$  
  & $0.309\pm0.535$ & $0.271\pm0.698$ & $0.165\pm 0.187$ \\
& $r_{\text{nextSNP,CHD}}$
& $(3.90\pm9.68)\cdot 10^{-3}$ & $(0.91\pm3.48)\cdot 10^{-3}$ & $(4.48\pm 7.13)\cdot 10^{-3}$\\
\hline
\end{tabular}
\caption{\baselineskip=0.5cm{
{\bf Probabilities inferred from the combined data sets excluding the extreme data sets.} Excluded: Georges e tal. \cite{georges_2003}, Bennet el al. \cite{bennet_2006} and Hou et al. \cite{hou_2009}. 
Similarly to Table~\ref{table:dyo}, 
to each hypothesis there correspond several rows consisting of: a) the parameters given by the maximum--likelihood values (one row in the case of $H_{0}$ and two rows in the case of $H_{1}$); b) the posterior probability for the occurrence of CHD (one row for each hypothesis); c) the predicted probabilities for the presence of the SNP (three rows for each hypothesis); and d) the probability ratio that measures the influence of CHD in the presence of the SNP, computed from the combined data of each phenotype (one row for each hypothesis). Column 1: The hypotheses. Column 2: The inferred quantities, as described above. Columns 3--5: The values of the inferred quantities for the combined ethnicity and CHD phenotype.}
}
\label{table:dyo_prob} 
\end{table}

\subsection{Sensitivity of the results}
\label{sec:sens}

To test the robustness of this meta--analysis, we conceive two tests of the sensitivity of the results, namely to low--significance data sets, to data sets with extreme results and to extreme data sets.  

To test the sensitivity of the results to low--significance data sets, we exclude the data sets with comparatively small sample sizes for the same CHD phenotype, namely the study by Elanhi et al. \cite{elahi_2008} and the study by Chen et al. \cite{chen_2001}, from the combination. We also exclude the studies with extreme results (i.e., the studies with the largest Bayes factor), namely the study in Dedoussis et al. \cite{dedoussis_2005}. 
We recompute both the Bayes factors (Table \ref{table:ena}) and the probabilities of CHD (Table \ref{table:dyo_improb}).
We observe that the Bayes factor in the new combination changes  
by $18\%,$  $-38\%$ and $24\%,$ respectively for the CS Caucasian, the MI Caucasian and the CS Asian population. 
The inferred parameters and probabilities vary 
by $-6$ to $6\%,$ $-5$ to $2\%,$ and $-1$ to $4\%,$ respectively for the CS Caucasian, the MI Caucasian and the CS Asian population.
The largest difference is observed for the CS Caucasian population due to the exclusion of the study by Elanhi et al. \cite{elahi_2008}. The exclusion of the study by Dedoussis et al. \cite{dedoussis_2005} from the MI Caucasian population causes predominantly negative differences.

To test the sensitivity of the results to extreme data sets, 
we exclude the data sets with comparatively large samples sizes for the same CHD phenotype, namely the study by Georges et al. \cite{georges_2003}, the study by Bennet el al. \cite{bennet_2006} and the study by Hou et al. \cite{hou_2009}, from the combination. These are also the studies with the smallest Bayes factor for each CHD phenotype. 
We recompute both the Bayes factors (Table \ref{table:ena}) and the probabilities of CHD (Table \ref{table:dyo_prob}).
We observe that the Bayes factor in the new combination changes  
by $3\%,$  $-19\%$ and $32\%,$ respectively for the CS Caucasian, the MI Caucasian and the CS Asian population.
The inferred parameters and probabilities vary 
by $-20$ to $-1\%,$ $5$ to $11\%,$ and $-26$ to $25\%,$ respectively for the CS Caucasian, the MI Caucasian and the CS Asian population.
The largest difference is observed for the CS Asian population due to the exclusion of the study by Hou et al. \cite{hou_2009}. The exclusion of the study by Georges et al. \cite{georges_2003} from the CS Caucasian population causes predominantly negative differences.

In both tests, the differences in the Bayes factor leave the result of the hypothesis testings unchanged, while the differences in the inferred parameters and probabilities also leave the conclusions unchanged.
We thus infer that this formalism is largely insensitive to a) low--significante data sets combined with data with extreme results, and to b) extreme data sets, which renders this formalism significantly robust. 

\section{Conclusions}
\label{sec:concl}

In this manuscript we investigated the correlation between the occurrence of CHD with the presence of the --308 TNF--$\alpha$ SNP from fifteen independent data sets on Caucasians for two CHD phenotypes and from five independent data sets on Asian for one CHD phenotype.
We showed how to combine independent data sets and to infer correlations using Bayesian analysis. 

Hypothesis testing on the combined data sets indicated that there is no evidence for a correlation between the occurrence of CHD and the presence of the SNP, either on Caucasians or on Asians. This result agrees with previous meta--analyses \cite{zhang_2011, chu_2013}. As a measure of an eventual correlation, we computed the conditional probability of CHD given the SNP, normalized to the probability that CHD occurs, finding that the presence of the SNP indicates the occurrence of CHD in of order 0.1\% of patients, i.e. in of order 0.1\% of the occurrence of CHD is concomitant with the presence of SNP. We also tested the sensitivity of the results by excluding selected data sets from the meta--analysis. We found changes of order $10\%,$ leaving the results unchanged and thus establishing this formalism as significantly robust. 

An interesting extension of this work for the sake of completion is the inclusion of studies referring to Africans and Indians which are currently too few to extract convincing results.\\

\noindent{\bf Acknowledgements} The author thanks E. Vourvouhaki and G. Tsiliki for useful discussions. 
The author is funded by Funda\c{c}\~ao para a Ci\^encia e a Tecnologia (FCT), Grant no. SFRH/BPD/65993/2009.


\begin{thebibliography}{99}

\bibitem{hansson_2005}
Hansson~GK, 
Inflammation, atherosclerosis and coronary heart disease,
N Engl J Med 352 (2005) 1685 


\bibitem{vassali_1992}
Vassali~P, The pathophysiology of tumor necrosis factors, 
Annu. Rev. Immunol 10 (1992) 411

\bibitem{plutzky_2001}
Plutzky~ J, 
Inflammatory pathways in atherosclerosis and acute coronary syndromes, 
Am J Cardiol 88 (2001) 10

\bibitem{vourvouhaki_2008}
Vourvouhaki~E and Dedoussis~GV, 
Cholesterol ester transfer protein: a therapeutic target in atherosclerosis?
Expert Opin. Ther. Targets 12 (2008) 937 



\bibitem{wilson_1992}
Wilson~AG, di Giovine~FS, Blakemore~AI and Duff~GW, 
Single base polymorphism in the humor tumour necrosis factor alpha (TNF alpha) gene detectable by NcoI restriction of PCR product, 
Hum Mol Genet 1 (1992)  353]
\bibitem{zhang_2011}
Zhang~HF, Xie~SL, Wang~JF, Chen~YX, Wang~Y and Huang~TC,
Tumor necrosis factor--alpha G--308A gene polymorphism and coronary hear disease susceptibility: an updated meta--analysis,
Thrombosis Research 127 (2011) 400

\bibitem{chu_2013}
Chu~H, Yang~J, Mi~S, Bhuyan~SS, Li~J, Zhong~L, Liu~S, Tao~Z, Li~J and Chen~H,
Tumor necrosis factor--alpha G-308 A polymorphism and risk of coronary heart disease and myocardial infarction: A case--control study and meta--analysis,
J Cardiovasc Dis Res, 3, (2013) 84

\bibitem{vourvouhaki_2009}
Vourvouhaki~E and Carvalho CS, 
A Bayesian approach to the probability of coronary heart disease subject to genetic risk factors, 
BioSystems 105 (2011) 181 (arXiv:0907.2043 [g-bio.QM])
\bibitem{vendrell_2003}
Vendrell~J, Fernandez-Real~J-M, Gutierrez~C., Zamora~A, Simon~I, Bardaji~A, Ricart~W and Richart~C,
A polymorphism in the promoter of the tumor necrosis factor--$\alpha$
gene (-308) is associated with coronary disease in type 2 diabetic patients, 
Atherosclerosis 167 (2003) 257 

\bibitem{elahi_2008}
Elahi~MM, Gilmour~A, Matata~BM and Mastana~SS,
A variant of position --308 of the Tumor necrosis factor alpha gene promoter and the risk of coronary heart disease,
Heart Lung Circ  17(1) (2008) 14 

\bibitem{allen_2001}
Allen~RA, Lee~EM, Roberts~DH, Park~BK and Pirmohamed~M,
Polymorphisms in the TNF--alpha and TNF--receptor genes in patients with coronary artery disease,
Eur J Clin Invest 31 (2001) 843

\bibitem{georges_2003}
Georges~JL, Rupprecht~HJ, Blankenberg~S, Poirier~O, Bickel~C, Hafner~G, Nicaud~V, Meyer~J, Cambien~F and Tiret~L,
Impact of pathogen burden in patients with coronary artery disease in relation to systemic inflammation and variation in genes encoding cytokines,
Am J Cardiol 92 (2003) 515

\bibitem{sbarsi_2008}
Sbarsi~I, Falcon~C, Boiocchi~C, Campo~I, Zorzetto~M, de Silvestri~A and Cuccia~M,
Inflammation and atherosclerosis: the role of TNF and TNH receptors polymorphisms in coronary artery disease,
Int J Immunopathol Pharmacol 20 (2007) 145

\bibitem{szalai_2002}
Szalai~C, F\"ust~G, Duba~J, Kramer~J, Romics~L, Proh\'aszka~Z and Cs\'asz\'ar~A,
Association of polymorphisms and allelic combinations in the tumour necrosis factor--alpha complement MHC region with coronary artery disease,
J Med Genet 39 (2002) 46
\bibitem{antonicelli_2005}
Antonicelli~R, Olivieri~F, Cavallone~L, Spazzafumo~L, Bonafe~M, Marchegiani~F, Cardelli~M, Galeazzi~R, Giovagnetti~S, Perna~GP and Franceschi~C,
Tumor necrosis factor--alpha gene --308G$\to$A polymorphism is associated with ST--elevation myocardial infarction and with high plasma levels of biochemical ischemia markers,
Coron Artery Dis 16 (2005) 489

\bibitem{bennet_2006}
Bennet~AM, van Maarle~MC, Hallqvist~J, Morgenstern~R, Frostegard~J, Wiman~B, Prince~JA and de Faire~U,
Association of TNF--alpha serum levels and TNFA--alpha promoter polymorphisms with risk of myocardial infarction,
Atherosclerosis 187 (2006) 408

\bibitem{dedoussis_2005}
Dedoussis~GV, Panagiotakos~DB, Vidra~NV, Louizou~E, Chrysohoou~C, Germanos~A, Mantas~Y, Tokmakidis~S, Pitsavos~C and Stefanadis~C,
Association between TNF--alpha --308G$\to$A polymorphism and the development of acute coronary syndromes in Greek subjects: the CARDIO2000-GENE Study,
Genet Med. 7(6) (2005) 411 

\bibitem{herrmann_1998}
Herrmann~SM, Ricard~S, Nicaud~V, Mallet~C, Arveiler~D, Evans~A, Ruidavets~JB, Luc~G, Bara~L, Parra~HJ, Poirier~O and Cambien~F,
Polymorphims of the tumor necrosis factor--alpha gene, coronary heart disease and obesity,
Eur J Clin Invest, 28 (1998) 59

\bibitem{koch_2001}
Koch~W, Kastrati~A, B\"{o}ttiger~C, Mehili~J, von Beckerath~N and Schomig~A,
Interleukin--10 and tumor necrosis factor gene polymorphisms and risks  of coronary artery disease and myocardial infarction,
Arterosclerosis 159 (2001) 137

\bibitem{padovani_2000}
Padovani~JC, Pazin--Filho~A, Simoes~MV, Marin--Neto JA, Zago MA, and Franco~RF,
Gene polymorphisms in the TNF locus and the risk of myocardial infarction,
Throm Res 100 (2000) 263

\bibitem{tobin_2004}
Tobin~MD, Braund~PS, Burton~PR, Thompson~JR, Steeds~R, Channer~K, Cheng~S, Lindpaintner~K and Samani~NJ,
ia multilocus case--control study, 
Eur Heart J 25 (2004) 459

\bibitem{tulyakova_2004}
Tulyakova~G, Nasibullin~T, Salmanov~A, Avzaletdinova~, Khusnutdinova~E, Zakirova~A and Mustafina~O,
Association of the --308 (G$\to$A) polymorphism of tumor necrosis factor--$\alpha$ with myocardial infarction and sudden cardiac death,
Balk J Med Genet 8 (2004) 31
\bibitem{chen_2001}
Chen~ZQ, Zheng~ZF, Ma~JF, Qiu~FY, Shi~Sl and Chen~H,
Plasma levels at TNF--$\alpha$ but not TNF--$\alpha$ G--308A gene polymorphism, is associated with coronary heart disease,
Chin J Geriatr Cardiaovas Cerebrovasc Dis 6 (2001) 279

\bibitem{hou_2009}
Hou~L, Huang~J, Lu~X, Wang~L, Fan~Z and Gu~D,
Polymorphisms of tumor necrosis factor--$\alpha$ gene and coronary hear disease in a Chinese Han population: interaction with cigarette smoking,
Thromb Res. 123 (2009) 822

\bibitem{li_2003}
Li~Y, Xu~P, Chen~H, Zhang~PA and Huang~CX,
Association between tumor necrosis factor--$\alpha$ G--308A gene polymorphism and coronary heart disease: a case--control study,
Chin J Geriatr 9 (2003) 568

\bibitem{liu_2009}
Liu~Y, Jin~W, Lu~L, Chen~QJ and Shen~WF,
Association between single nucleotide polymorphism in the promoter of tumor necrosis factor--$\alpha$ gene and coronary heart disease,
J Diagn Concepts Pract 5 (2009) 506

\bibitem{shun_2009}
Shun~SY, Zeng~XQ, Qi~AM, Fan~WH and Zhang~JC,
Association of tumor necrosis factor--$\alpha$ gene polymorphisms in the promoter region with chronic periodontitis and coronary heart disease,
J Clin Stomatol 5 (2009) 279


\bibitem{walker_2008}
Walker~E, Hernandez~AV and Kattan~MW,
Meta--analysis: Its strengths and limitations,
Cleveland Clinic J Medicine, 75, 6  (2008) 431

\bibitem{mackay}
MacKay~DJC, 
Information Theory, Inference, and Learning Algorithms,
Cambridge University Press, 2003

\bibitem{jeffrey}
Kass~RE and Raftery~AE,
Bayes Factors, 
J American Statistical Association 90, 430 (1995) 773


\bibitem{pcr_error}
Sharifian~H,
Errors induced during PCR amplification, 
Master Thesis ETH Zurich (2010) http://dx.doi.org/10.3929/ethz-a-006088024

\bibitem{esc_cs}
Hamm~CW, Bassand~JP, Agewall~S, Bax~J, Boersma~E, Bueno~H, Caso~P, Dudek~D, Gielen~S, Huber~K et al.,
ESC Guidelines for the management of acute coronary syndromes in patients presenting without persistent ST--segment elevation,
Eur Heart J 32, (2011) 2999 

\bibitem{esc_mi}
Steg~PG, James~SK, Atar~D, Badano~L, Bl\"omstrom-Lundqvist C, Borger~MA, Di Mario~C, Dickstein~K, Ducrocq~G, Fernandez--Aviles~F et al.
ESC Guidelines for the management of acute myocardial infarction in patients presenting with ST--segment elevation,
Eur Heart J 33, (2012) 2569 


\end{thebibliography}
\end{document}